\documentclass[11pt,a4paper]{article}
\pdfoutput=1
\usepackage{jcappub}
\usepackage[colorlinks=true,linkcolor=blue,citecolor=blue,urlcolor=blue]{hyperref}
\usepackage[sort&compress]{natbib}
\usepackage{amsmath,amssymb,amsthm,cleveref,aas_macros,siunitx,bm}
\usepackage{xcolor}

\newcommand\OmegaPBH{\Omega_{\mathrm{PBH}}}
\newcommand\psimono{\psi_{\mathrm{mono}}}
\newcommand\fmono{f_{\mathrm{mono}}}
\newcommand\fmaxmono{f_{\mathrm{max}}}
\newcommand\Aobs{A_{\mathrm{obs}}}
\newcommand\fmax[1]{f_{\mathrm{max},#1}}
\newcommand\ftotmono{\fmono}
\newcommand\fpbh{f_{\mathrm{PBH}}}
\newcommand\fbound{f_{\mathrm{max},\mathrm{all}}}
\newcommand\fgw{f_{\mathrm{max},\mathrm{GW}}}
\newcommand\du{\mathrm{d}}
\newcommand\dd{\,\du}
\newcommand\norm[1]{\left\|#1\right\|}
\newcommand{\bb}[1]{\bm{\mathrm{#1}}}
\newcommand{\cvec}{\bb{\mathcal C}}
\newcommand{\convg}{\operatorname{conv}(\bb g)}
\newcommand{\cset}[1]{$\bm{\mathrm{\mathsf{#1}}}$}
\newcommand{\ngw}{\nu_{\mathrm{GW}}}

\title{The Maximal-Density Mass Function for Primordial Black Hole Dark Matter}
\author{Benjamin V. Lehmann,}
\emailAdd{blehmann@ucsc.edu}
\author{Stefano Profumo}
\emailAdd{profumo@ucsc.edu}
\author{and Jackson Yant}
\emailAdd{jyant@ucsc.edu}
\affiliation{Department of Physics, University of California Santa Cruz,\\
1156 High St., Santa Cruz, CA 95064, USA}
\affiliation{Santa Cruz Institute for Particle Physics,\\
1156 High St., Santa Cruz, CA 95064, USA}
\abstract{
The advent of gravitational wave astronomy has rekindled interest in primordial black holes (PBH) as a dark matter candidate. As there are many different observational probes of the PBH density across different masses, constraints on PBH models are dependent on the functional form of the PBH mass function. This complicates general statements about the mass functions allowed by current data, and, in particular, about the maximum total density of PBH. Numerical studies suggest that some forms of extended mass functions face tighter constraints than monochromatic mass functions, but they do not preclude the existence of a functional form for which constraints are relaxed. We use analytical arguments to show that the mass function which maximizes the fraction of the matter density in PBH subject to all constraints is a finite linear combination of monochromatic mass functions. We explicitly compute the maximum fraction of dark matter in PBH for different combinations of current constraints, allowing for total freedom of the mass function. Our framework elucidates the dependence of the maximum PBH density on the form of observational constraints, and we discuss the implications of current and future constraints for the viability of the PBH dark matter paradigm.
}

\begin{document}
\maketitle

\section{Introduction}
\label{sec:introduction}
The possibility that density fluctuations in the early universe collapsed into primordial black holes (PBH) has been studied for several decades \citep{Carr1974}. Apart from their potential utility as a probe of the primordial universe, PBH are an excellent candidate for cosmological dark matter, as sufficiently large black holes are stable and dynamically cold. Further, with simple formation mechanisms, they can be produced with a cosmological density matching the observed density of dark matter. 

If PBH account for a significant fraction of dark matter, it is possible that observed gravitational wave signals have a primordial origin. Direct observations of binary black hole mergers thus far all involve black holes with masses of several times $10M_\odot$ \cite{GW1,GW2,GW3,GW4,GW5}, in a range where microlensing constraints on the abundance of compact objects are ineffective. The observed merger rate is compatible with PBH as dark matter, and other constraints historically applied in the LIGO mass range are subject to large astrophysical uncertainties \cite{Bird2016,Ali-Haimoud:2016mbv}. This has led to renewed interest in primordial production mechanisms, and it remains possible that PBH in this mass window account for much or all of dark matter \citep{Garcia-Bellido2017}.

However, depending on the formation mechanism, PBH may exist today with masses as small as $10^{-16}M_\odot$, or as large as those of supermassive black holes. Thus, constraining the total density contained in PBH requires the combination of constraints that span this vast range of mass scales. Such observables include microlensing surveys \cite{hsc_Niikura2017,kepler_Griest2014,constraint_EROS,MACHO_Alcock2001}, CMB data \cite{Planck_Ali-Haimoud2017}, and the statistics of wide binaries \cite{WB_Monroy-Rodriguez2014}. In general, constraints from these observables have been computed under the assumption that all PBH have the same mass. The corresponding mass functions, comprising a single Dirac delta, are said to be \emph{monochromatic}. However, as realistic production mechanisms necessarily result in an extended (non-monochromatic) mass function, it is essential to correctly combine constraints across all masses.

This problem has recently been studied by several authors \cite{Kuhnel2017,Carr2017,Bellomo2017}. In general, the constraints depend non-trivially on the functional form of the mass function, and statements about the implications of constraints for properties of the PBH population can be difficult to generalize. In particular, the total fraction $\fpbh$ of dark matter that may be accounted for by PBH varies with the form of the mass function, so $\fpbh=1$ is ruled out for some forms of the mass function, and allowed for others. This has led to confusion regarding the observational viability of the PBH dark matter scenario, and while prior work has established procedures for comparing specific extended mass functions with observables, general statements regarding the allowed total fraction of dark matter in PBH are lacking.

Depending on the set of constraints considered, observational data may or may not already rule out $\fpbh=1$ for monochromatic mass functions. Since the many constraints span a wide mass range, and since several do not overlap significantly, some authors have argued that broadening the mass function might relax constraints on PBH \cite{Inomata:2017okj,Clesse:2015wea}, possibly allowing for $\fpbh=1$ even if that possibility were excluded by constraints for monochromatic mass functions. However, \cite{Carr2017,Bellomo2017} have evaluated the constraints \emph{numerically} for several forms of extended mass functions, and found that extended mass functions are typically subject to stronger constraints than monochromatic mass functions.

These findings motivate the question we now pose: what is the theoretical maximum density of PBH permitted by constraints for a fully general mass function? Our goal is ultimately to clarify the observational status of PBH dark matter, and to understand the circumstances under which extending the mass function can relax constraints. We also seek a procedure which is flexible and simple enough to allow us to compare results for different sets of constraints, and to elucidate the dependence of the maximal density on the form of the constraints themselves. To that end, we derive the form of the mass function which optimizes the density subject to all observational constraints combined. This allows us to obtain a general bound on the density of PBH with minimal numerical computation, independently of the true form of the PBH mass function. Note that we do not propose a new prescription for the evaluation of constraints for a given extended mass function. Rather, we maximize the PBH density subject to constraints as evaluated using existing methods from the literature. The maximal-density mass functions we derive then provide insights into the overall impact of each individual observable.

This paper is organized as follows. In~\cref{sec:constraints}, we establish conventions and notations, and review the application of constraints from the monochromatic case to extended mass functions. In~\cref{sec:optimization}, we present a pedagogical derivation of our main results regarding the maximum density of PBH, and we apply them to current data. We consider the impact of gravitational wave constraints separately in~\cref{sec:gravitational-waves}. We discuss these results in~\cref{sec:discussion} and conclude in~\cref{sec:conclusions}. Finally, in~\cref{sec:numerical}, we validate our analytical results with direct numerical techniques.

\section{Interpreting constraints for extended mass functions}
\label{sec:constraints}
\subsection{The interpretation problem}
Applying observational constraints to generic extended mass functions is non-trivial. It is not sufficient to check that the mass function does not intersect constraint curves, as experiments are typically sensitive to the \emph{integral} of the mass function in each of a set of mass bins. Thus, most constraints are only trivial to interpret for monochromatic mass functions, i.e., mass functions of the form
\begin{equation}
\psimono\left(M_0,f_0;M\right)\equiv f_0\,\delta\left(M-M_0\right)
\end{equation}
whose integrals are non-zero in only one bin. In this case, an observational constraint curve $\fmaxmono(M)$ imposes the requirement that $f_0<\fmaxmono(M_0)$. Transforming such constraints to the parameter space of a more general extended mass function involves summing contributions to observables from all mass bins. Multiple prescriptions for this procedure have been used in the literature.

The earliest systematic treatment of constraints for extended mass functions is due to~\cite{Carr2016}. They divide the mass range into $N$ bins $I_1,\dotsc,I_N$, approximating the constraint functions as step functions on these bins. Within each bin, only the strongest constraint function $\fmaxmono(M)$ is considered. A mass function $\psi$ is excluded if
\begin{equation}
\label{eq:discrete-constraint}
\int_{I_k}\du M\,\psi(M)>\max_{M\in I_k}\fmaxmono(M)
\end{equation}
for any $k$.
This prescription is used by~\cite{Kuhnel2017} to numerically transform observational constraints to the parameter space of a lognormal mass function. Their findings suggest that broadening the mass function does not generally relax constraints. However, as~\cite{Kuhnel2017} treat the problem computationally, it is difficult to determine the relevance to their results of any particular constraint, or of the lognormal form of the trial mass functions. This provided partial motivation for the analysis of~\cite{Carr2017}, who obtain similar numerical results for several additional constraints and forms of the mass function. Further, \cite{Carr2017} derive a more rigorous prescription for transforming observational constraints to general extended mass functions. We review their derivation in~\cref{sec:constraint-prescription}.

Similar questions motivate the recent analysis of~\cite{Bellomo2017}. Rather than develop a prescription for translating constraints for monochromatic mass functions to suit a given extended mass function, the authors develop a prescription for converting the extended mass function into a set of monochromatic mass functions, each accounting for the contribution of the PBH population to one observable. The extended mass function is then subject to each constraint as it applies to the corresponding monochromatic mass function. This approach is used to constrain the parameter spaces of lognormal and power law mass functions, with results similar to those of~\cite{Kuhnel2017} and~\cite{Carr2017}.

Our methods bear some similarities to~\cite{Bellomo2017}, in that we also find it sufficient to work with sets of monochromatic mass functions. However, the monochromatic mass functions we consider have a different interpretation, as discussed in~\cref{sec:general}. Our goal is not to place constraints on any specific extended mass function, but rather to place bounds on PBH dark matter while allowing complete freedom in the mass function. Thus, our formalism is structured around the maximization problem, and we use our results to study both the current status of the PBH dark matter paradigm and the potential impact of future observables.

\subsection{Constraint prescription}
\label{sec:constraint-prescription}
In this work, we seek a general result for the maximum allowed fraction of dark matter in PBH, independent of the form of the mass function, and in a form that elucidates the relevance of each observable. As such, it is necessary that we adopt a prescription for constraining a given mass function that allows for multiple simultaneous constraining observables, a requirement most naturally satisfied by that of~\cite{Carr2017}. Their prescription is thus the basis for our analytical work. We numerically confirm that similar results are obtained under the prescriptions of~\cite{Bellomo2017} and~\cite{Carr2016} (see~\cref{sec:prescription-sensitivity}).

We follow~\cite{Carr2017} to convert constraints for monochromatic mass functions to constraints for extended mass functions. We denote the mass function by $\psi$ and adopt their normalization and conventions, such that 
\begin{equation}
\psi\propto M\frac{\du n}{\du M},
\qquad
\int\du M\,\psi(M)=\frac{\OmegaPBH}{\Omega_{\mathrm{DM}}}\equiv \fpbh
\end{equation}
where $n$ is the number density of PBH at fixed mass. Most observables that can constrain primordial black holes are determined by the properties of single black holes, with no need to consider relationships between them. In such a case, an observable quantity $A$ receives a linear combination of contributions from each mass bin, and the contribution from black holes of mass $M$ is proportional to $\psi(M)$. As such, the observable can be written as a functional of $\psi$ in the form
\begin{equation}\label{eq:firstorder}
A[\psi]=A_0+\int\du M\,\psi(M)K_1(M)
.
\end{equation}

We note in passing that there are some observables for which relationships between black holes are significant. For example, gravitational wave observations of mergers are dependent on the properties of pairs of black holes, and so one must combine contributions from pairs of mass bins. In the simplest case, where the contributions scale linearly with number in each mass bin, such an observable can clearly be written in the form
\begin{equation}
\label{eq:psi-observable}
A[\psi]=A_0+\int\du M\,\psi(M)K_1(M)+\int\du M\dd M^\prime\,\psi(M)\psi(M^\prime)K_2(M,M^\prime)
\end{equation}
and one can always express a generic observable by including higher-order terms of this form. Note that higher-order terms also account for non-linear dependence of $A$ on $\psi$ at fixed mass. For example, an observable which scales as $\psi(M)^2$ can be expressed exactly at second order by setting $K_2(M,M^\prime)\propto\delta(M-M^\prime)$.

We study the potential impact of gravitational wave observations in~\cref{sec:gravitational-waves}.
All of the other constraints that we consider in this work are of the simplest kind, and we will find~\cref{eq:firstorder} sufficient. In this case, it is straightforward to relate constraints for a monochromatic mass function to constraints for a generic mass function, and we briefly review the argument given in~\cite{Carr2017}. Let $\psimono(M_0;M)\equiv \fmaxmono(M_0)\,\delta(M-M_0)$, where $\fmaxmono(M_0)$ is the largest coefficient allowed by constraints for a mass function of this form. If we take $\psi(M)=\psimono(M_0;M)$ in~\cref{eq:psi-observable}, we obtain
\begin{equation}
\label{eq:observable-kernel}
K_1(M_0)=\frac{A[\psimono]-A_0}{\fmaxmono(M_0)}
\end{equation}
Suppose that the difference $A[\psi]-A_0$ is observable with the desired significance when $A[\psi]$ crosses a threshold value $\Aobs$. Then $A[\psimono]=\Aobs$ by definition of $\fmaxmono$, so~\cref{eq:observable-kernel} gives $K_1(M)$ independent of $\psi$. Substituting for $K_1(M)$ in~\cref{eq:psi-observable} while leaving $\psi$ generic gives the condition
\begin{equation}\label{eq:constraint-form}
\mathcal C[\psi]\equiv\int\du M\frac{\psi(M)}{\fmaxmono(M)}\leq1
.
\end{equation}
This expresses the constraint on a mass function $\psi(M)$ when the constraint for a monochromatic mass function is $\int\du M\,\psimono(M_0;M)\leq\fmaxmono(M_0)$.

\section{The optimal mass function}
\label{sec:optimization}
\subsection{Single-constraint case}
\label{sec:single-constraint-case}
For pedagogical purposes, we first consider the case of a single constraining observable. For such situations, when all observables can be expressed in the form of~\cref{eq:firstorder}, the constraint on the mass function has the form $\mathcal C[\psi]\leq1$, with $\mathcal C[\psi]$ as defined in~\cref{eq:constraint-form}. The problem is then to maximize $\int\du M\,\psi(M)$ subject to this constraint. The optimal mass function saturates the constraint, so it suffices to require $\mathcal C[\psi]=1$.

Naively, this problem looks as though it can be solved using the method of Lagrange multipliers, by finding stationary points of the functional
\begin{equation}
\mathcal S[\psi,\lambda]=\int\du M\,\left(\psi(M)-\lambda\frac{\psi(M)}{\fmaxmono(M)}\right)
.
\end{equation}
However, the Euler-Lagrange equation in $\psi$ admits no non-trivial solutions. This is because $\int\du M\,\psi(M)$ can be made arbitrarily large, even subject to $\mathcal C[\psi]=1$, unless $\psi(M)>0$ is imposed. Positivity can be imposed by setting $\psi=\phi^*\phi$ and performing an unconstrained optimization in $\phi$, but the corresponding Euler-Lagrange equation leads to the condition that $\phi$ is, at every point, either zero or non-analytic.

The variational approach does not generalize to the case of multiple constraints, so we do not pursue it any further. Rather, we observe that since $\mathcal C[\psi]$ is linear, we have $\mathcal C\bigl[\mathcal C[\psi]^{-1}\psi\bigr]=1$. Thus, we can impose $\mathcal C[\psi]=1$ by rescaling $\psi\to\mathcal C[\psi]^{-1}\psi$, and then the problem is to maximize the functional
\begin{equation}
\mathcal M[\psi]\equiv\int\du M\left(\mathcal C[\psi]^{-1}\psi(M)\right)=\frac{\int\du M\,\psi(M)}{\int\du M\,\frac{\psi(M)}{\fmaxmono(M)}}
\end{equation}
subject only to positivity. We call $\mathcal M[\psi]$ the \emph{normalized mass} of $\psi$.

It is now simple to show that $\mathcal M[\psi]$ is maximized by taking $\psi$ to be a monochromatic mass function. Let $M_{\mathrm{max}}\equiv\operatorname{argmax}\fmaxmono(M)$ and $\fmono\equiv\fmaxmono(M_{\mathrm{max}})$, and define
\begin{equation}
\psi_0(M)\equiv\fmono\ \delta(M-M_{\mathrm{max}})
\end{equation}
so that $\psi_0(M)$ is the monochromatic mass function which maximizes the PBH density, and $\fmono$ is the maximum PBH density allowed for a monochromatic mass function. Choose any mass function $\psi\equiv\psi_0+\delta\psi$. Since $\psi_0$ vanishes everywhere except for $M_{\mathrm{max}}$, positivity of $\psi$ requires that $\delta\psi(M)\geq0$ for all $M\neq M_{\mathrm{max}}$. Then we have
\begin{equation}
\mathcal M[\psi]=\frac{\int\du M\bigl[\psi_0(M)+\delta\psi(M)\bigr]}{\int\du M\bigl[\psi_0(M)/\fmaxmono(M)+\delta\psi(M)/\fmaxmono(M)\bigr]}
.
\end{equation}
Since $\psi_0$ saturates the constraint of~\cref{eq:constraint-form}, we must have $\int\du M\left[\psi_0(M)/\fmaxmono(M)\right]=1$ and $\int\du M\,\psi_0(M)=\fmono$, so we write
\begin{equation}
\mathcal M[\psi]=\frac{\fmono+\int\du M\,\delta\psi(M)}{1+\int\du M\,\delta\psi(M)/\fmaxmono(M)}
\end{equation}
but $\fmaxmono(M)\leq\fmono$ by definition, so we have
\begin{equation}
\mathcal M[\psi]=\frac{\fmono+\int\du M\,\delta\psi(M)}{1+\int\du M\,\delta\psi(M)/\fmaxmono(M)}
\leq\frac{\fmono+\int\du M\,\delta\psi(M)}{1+\int\du M\,\delta\psi(M)/\fmono}=\fmono
.
\end{equation}
Thus we have shown that $\mathcal M[\psi]\leq\fmono\equiv\mathcal M[\psi_0]$, so no functional form allows a higher total PBH density than does the Dirac delta. In particular, for fixed PBH density, we conclude that an extended mass function is always more strongly constrained than the optimal monochromatic mass function. While this will not hold for the case of multiple constraints, it remains an excellent approximation if the constraints are weakest by far in a mass range where a single observable dominates.

\subsection{Combining constraints}
Realistically, the single-constraint case is too simplistic. In general, a mass function is ruled out on the basis of a $\chi^2$ test statistic. If PBH are constrained by multiple observables $A_j$, then the test statistic is found by adding the individual $\chi^2$ statistics in quadrature. That is,
\begin{equation}
\chi^2[\psi]=\sum_{j=1}^N\chi_j^2=\sum_{j=1}^N\left(\frac{A_j[\psi]-A_{\mathrm{obs},j}}{\sigma_j}\right)^2
.
\end{equation}
To fail to reject $\psi$ at some significance level requires that $\chi^2[\psi]\leq\gamma^2$ for some threshold value $\gamma^2$, i.e.,
\begin{equation}
\sum_{j=1}^N\left(\int\du M\,\psi(M)\frac{K_{1,j}(M)}{\gamma\sigma_j}\right)^2\leq1
.
\end{equation}
If we set $N=1$, this reduces to
\begin{equation}
\int\du M\,\psi(M)\frac{K_{1,1}(M)}{\gamma\sigma_1}\leq1
\end{equation}
so matching with~\cref{eq:constraint-form} gives $K_{1,j}(M)/(\gamma\sigma_j)=1/\fmax{j}(M)$, where $\fmax{j}(M)$ is the analogue of $\fmaxmono(M)$ for the $j$th constraint alone. For general $N$, \cite{Carr2017} show that the constraint takes the form
\begin{equation}
\label{eq:general-constraint}
\sum_{j=1}^N\left(\int\du M\,\frac{\psi(M)}{\fmax{j}(M)}\right)^2\leq1
.
\end{equation}
Since the individual constraints are added in quadrature, the argument applied to the single-constraint case does not extend to the case of multiple constraints, and indeed, there are cases in which the density is not maximized by a monochromatic mass function. However, we will show that the maximizer is in general a linear combination of $N$ monochromatic mass functions.

\subsection{The general problem}
\label{sec:general}
For the case of several constraining observables, one has $N$ constraint functions denoted by $\fmax{1},\dotsc,\fmax{N}$. For brevity, we define $g_j(M)\equiv1/\fmax{j}(M)$, and by analogy with~\cref{eq:constraint-form}, we define
\begin{equation}
\mathcal C_j[\psi]\equiv\int\du M\,\psi(M)\,g_j(M)
.
\end{equation}
Then the problem is to find $\psi$ to maximize
\begin{equation}\label{eq:multi-normalized-mass}
\mathcal M[\psi]\equiv\frac{\int\du M\,\psi(M)}{\left(\sum_{j=1}^N\mathcal C_j[\psi]^2\right)^{1/2}}
=\frac{\int\du M\,\psi(M)}{\norm{\cvec[\psi]}}
\end{equation}
where $\cvec[\psi]$ denotes the vector with components $\mathcal C_j[\psi]$. We define $\fbound=\max\mathcal M[\psi]$.

Since rescaling $\psi$ does not change $\mathcal M[\psi]$, we can always set $\int\du M\,\psi(M)=1$, and then the problem is equivalent to minimizing $\norm{\cvec[\psi]}$ subject to this constraint. For convenience, we cast the integral in discrete form, writing
\begin{equation}
\norm{\cvec[\psi_Q]}^2=\sum_{j=1}^N\left(\sum_{k=1}^Qa_kg_j(M_k)\right)^2=\norm{\sum_{k=1}^Qa_k\bb g(M_k)}^2
\end{equation}
where $Q$ is not restricted to be finite.
Thus, the problem is to minimize the norm of a sum of $a_k\bb g(M_k)$ for some $\{M_k\}_{k=1,\dotsc,Q}$, subject to our normalization condition, which now takes the form $\sum_{k=1}^Qa_k=1$. Geometrically, this is the same as minimizing the norm over the convex hull of the $\bb g(M)$, i.e., to compute
\begin{equation}
\min\left\{
\norm{\bb x}
\;\middle|\;
\bb x\in\operatorname{conv}\left\{
		\bb g(M)\;\middle|\;M\in U
	\right\}
\right\}
\end{equation}
where $U$ is the mass range under consideration.
We henceforth denote $\operatorname{conv}\left\{
		\bb g(M)\;\middle|\;M\in U
	\right\}$ by $\convg$. Since the minimizer is the projection of the origin onto a convex set, it is unique in the sense that any optimal mass function $\psi$ must have the same $\cvec[\psi]$. This does not require that the minimizing mass function is itself unique.

Such a geometric formulation simplifies the interpretation of the problem. In particular, the result for the case of a single constraint is now immediate: the convex hull is 1-dimensional, so the point with minimum norm is simply the minimum value of $g(M)$. The corresponding mass function is monochromatic, with a peak at $\operatorname{argmin}g(M)$. It is also clear that the monochromatic mass function is not generally the minimizer of the norm in the case of multiple constraints: we have no guarantee that $\norm{\bb g(M)}$ attains the minimum of the norm on $\convg$ for any single $M$.

Still, minimizing the norm over the convex hull of a discretization of $\bb g(M)$ is a simple computational problem, and it is easy to validate the result. We find an optimal mass function in three steps:
\begin{enumerate}
\item\label{item:1} Choose a discretization of $\bb g(M)$ of the form $G=\left\{\bb g(M_1),\dotsc,\bb g(M_R)\right\}$. We choose the $M_k$ using adaptive sampling to capture features of the constraint functions as precisely as possible. The convex hull of $G$ is now a polytope $A$.
\item Find the point $p_{\mathrm{min}}\in A$ with minimum norm. We implement the algorithm of~\cite{Wolfe1976}, which requires only the extreme points of $A$ as inputs. To avoid computing the convex hull in a high-dimensional space, we supply all of the points of $G$, of which the extreme points of $A$ form a subset. The algorithm determines the facet $S$ of $A$ which contains $p_{\mathrm{min}}$, and gives the barycentric coordinates of $p_{\mathrm{min}}$ in $S$ as a vector $\bb w$.
\item Define a mass function
\begin{equation}
\psi_{\mathrm{opt}}(M)=\sum_{k=1}^{\left|\bb w\right|}w_k\delta\left(M-M_k\right)
\end{equation}
where $\bb g(M_k)$ is the $k$th point of $S$. Note that $\bb g(M_k)\in G$ for each $M_k$ since $S\subset A$.
\end{enumerate}
Observe that $\bb{\mathcal C}[\psi_{\mathrm{opt}}]=\sum_{k=1}^{\left|\bb w\right|}w_k\bb g\left(M_k\right)\equiv p_{\mathrm{min}}$. Thus, $\psi_{\mathrm{opt}}$ is a mass function which attains the maximum total dark matter fraction. In particular, for any mass function $\psi$, we have $\mathcal M[\psi]\leq\mathcal M[\psi_{\mathrm{opt}}]=\norm{p_{\mathrm{min}}}^{-1}$, so $\fbound=\norm{p_{\mathrm{min}}}^{-1}$ is an upper bound on the fraction of dark matter in PBH irrespective of the functional form of the mass function. We will refer to $\psi_{\mathrm{opt}}$ as the \emph{semi-analytical optimum} mass function.

We can now explain geometrically why the maximizing mass function is a linear combination of no more than $N$ monochromatic mass functions. Observe that for any $\bb g(M)$, the minimum of the norm must lie on the boundary of the convex hull $\convg$, and since $\bb g(M_k)\in\mathbb R^N$, this boundary has dimension at most $N-1$. One can construct an arbitrarily refined triangulation of this boundary formed from $(N-1)$-simplices, each with $N$ points of $G$ as vertices. The minimizer of the norm is a linear combination of these vertices, each of which is one of the original $\bb g(M_k)$, corresponding to a monochromatic mass function. We emphasize that at no step do we impose that the optimal mass function is a discrete linear combination of a finite number of monochromatic mass functions. This is a consequence of the fact that the optimum corresponds to a point in an $(N-1)$-simplex, meaning that this mass function lies in a space which is spanned by at most $N$ monochromatic mass functions.

Our method is deceptively similar to the procedure of~\cite{Bellomo2017}, in that we also work with sets of monochromatic mass functions. However, the monochromatic mass functions considered in that work are used only to study the consequences of a given extended mass function for a given observable. The sum of the effective monochromatic mass functions corresponding to each observable does not generally give a single mass function with equivalent consequences for all observables combined. This is appropriate for the purposes of~\cite{Bellomo2017} because they investigate which constraints are most effective for mass functions of a fixed functional form.

In this work, the mass functions we derive maximize the density of PBH with respect to all constraints simultaneously. Since the constraints are statistical in nature, the combination of multiple independent constraints at a single mass is stronger than any one of them individually. We follow~\cite{Carr2017} in treating constraints simultaneously, and our resulting semi-analytical optima are indeed sums of monochromatic mass functions. This approach is necessary for our purposes because we investigate which constraints and mass ranges are most significant for overall constraints on PBH dark matter, irrespective of the functional form of the mass function.

\subsection{Results}
\label{sec:results}
We perform the maximization explicitly for several sets of constraints. Set \cset{A} includes robust constraints from evaporation~\citep{evap_Carr2010}; GRB lensing~\citep{fl_Barnacka2012}; microlensing from HSC~\citep{hsc_Niikura2017}, Kepler~\citep{kepler_Griest2014}, EROS~\citep{constraint_EROS}, and MACHO~\citep{MACHO_Alcock2001}; and CMB limits from Planck~\citep{Planck_Ali-Haimoud2017}. Set \cset{B} includes dynamical constraints from Segue I~\citep{Seg_Koushiappas2017}, Eridanus II~\citep{Eri_Brandt2016}, and non-disruption of wide binaries~\citep{WB_Monroy-Rodriguez2014}. Set \cset{C} includes a constraint from white dwarf explosions~\citep{constraint_WD}, a constraint from neutron star capture~\citep{constraint_NS} and a recently claimed constraint from SNe lensing in the LIGO window~\citep{constraint_SNe}. The constraints from evaporation and from Planck in \cset{A} have been estimated differently in the literature, with important consequences for our analysis. Set \cset{A} itself contains relatively non-restrictive estimates of these constraints. We incorporate more stringent versions (see~\cref{sec:discussion}) of these constraints in a set \cset{\bar A}, which is otherwise identical to \cset{A}.

\begin{figure*}\centering
\includegraphics[width=0.50\textwidth]{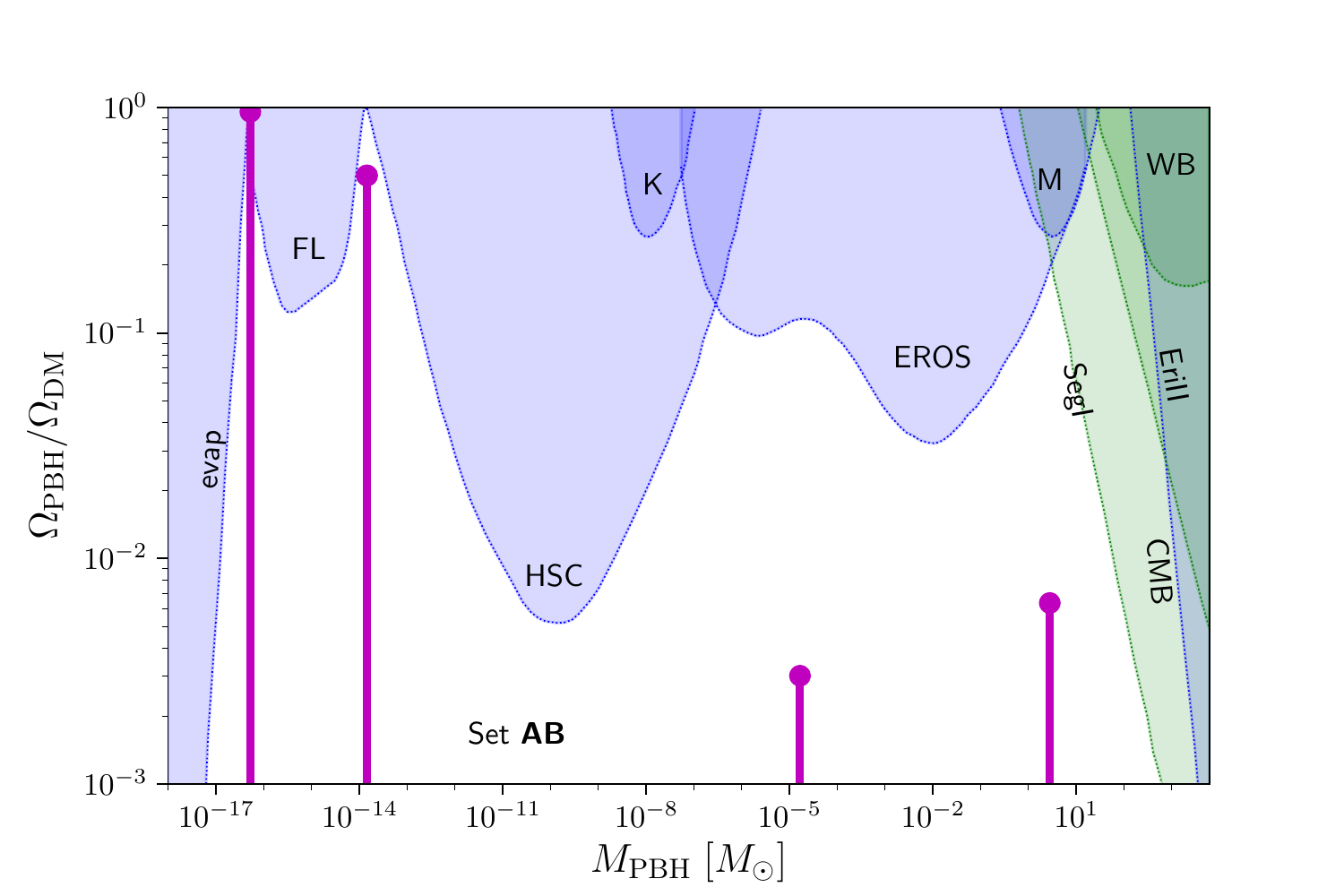}
\hspace{-0.04\textwidth}
\includegraphics[width=0.50\textwidth]{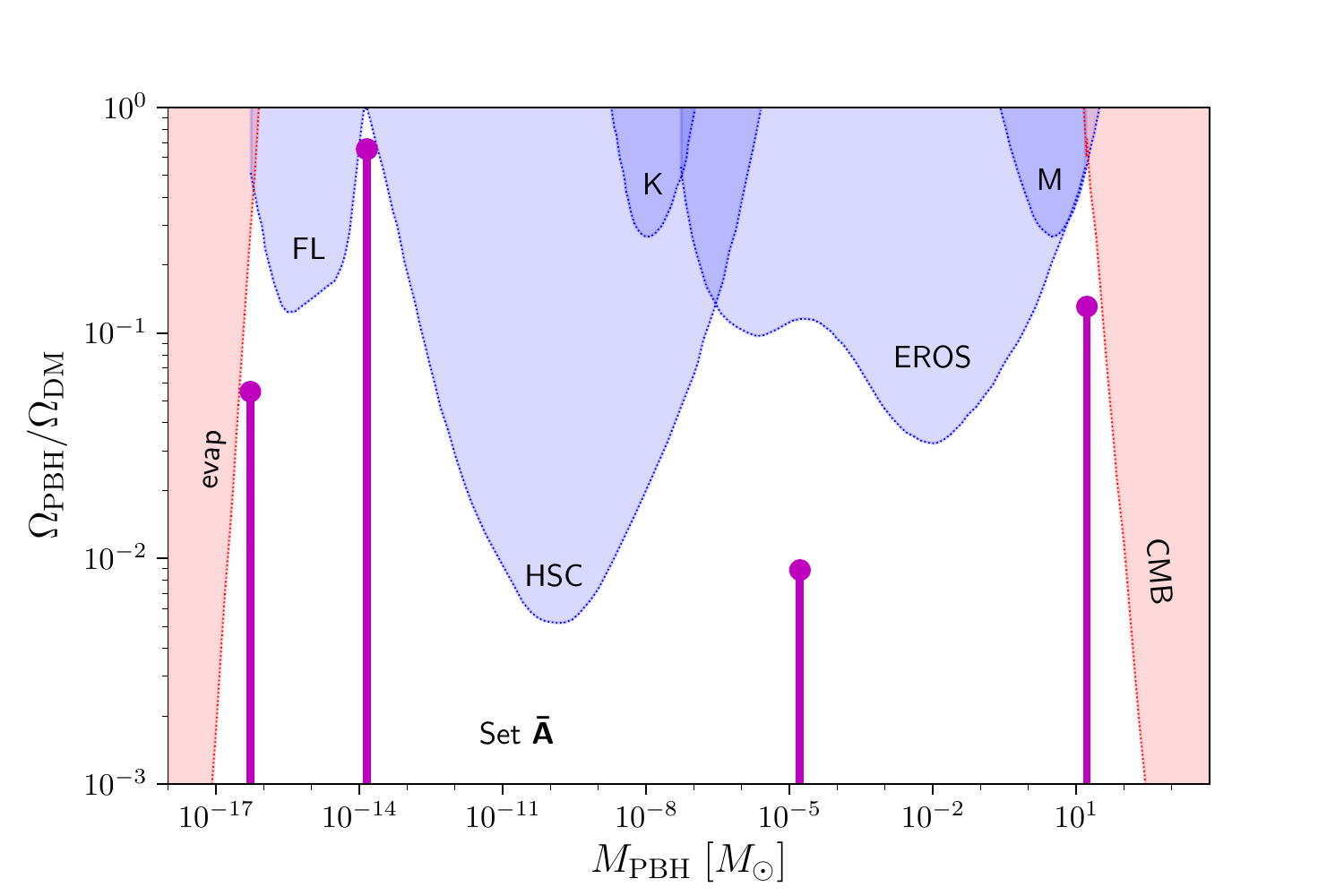}
\\
\includegraphics[width=0.50\textwidth]{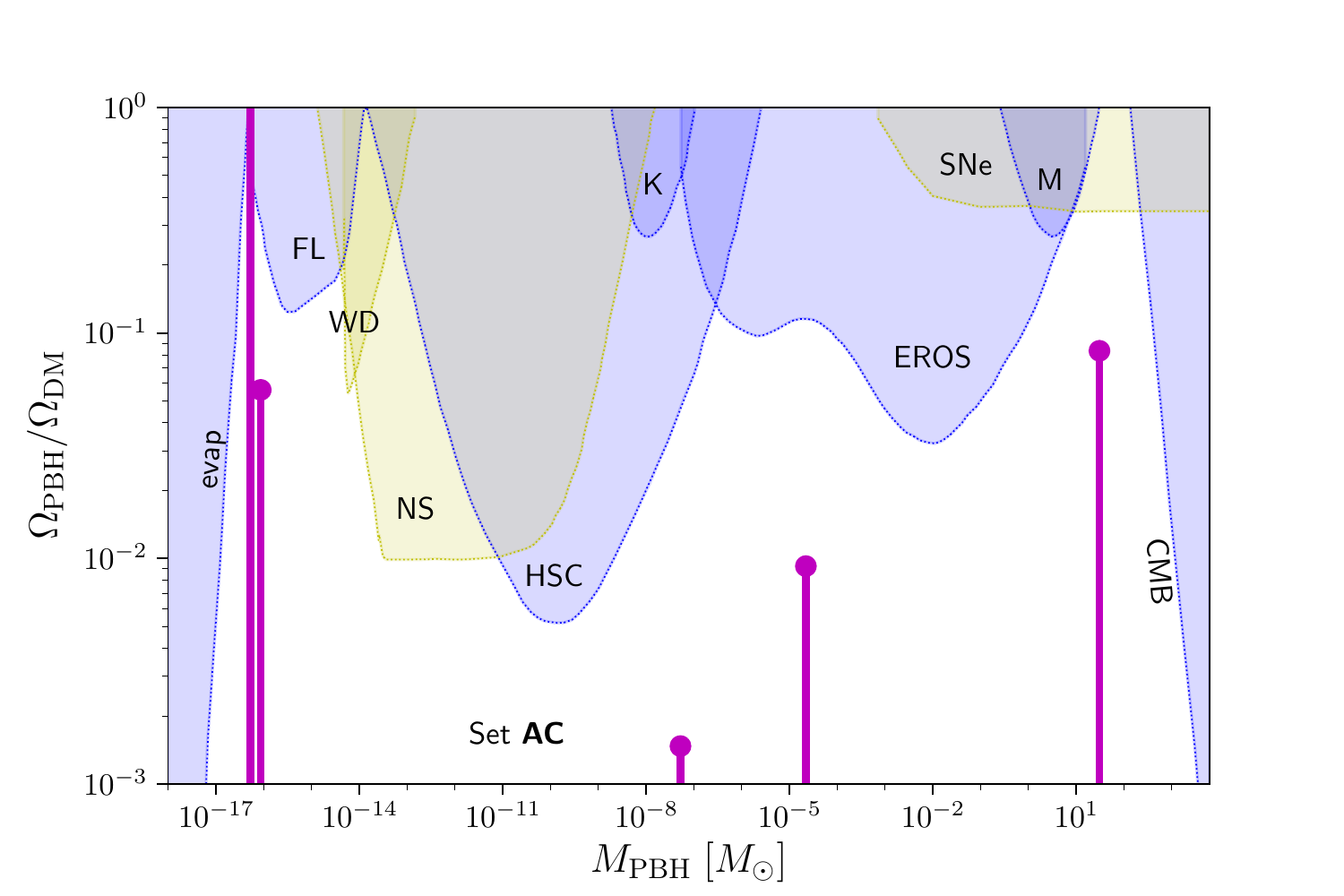}
\hspace{-0.04\textwidth}
\includegraphics[width=0.50\textwidth]{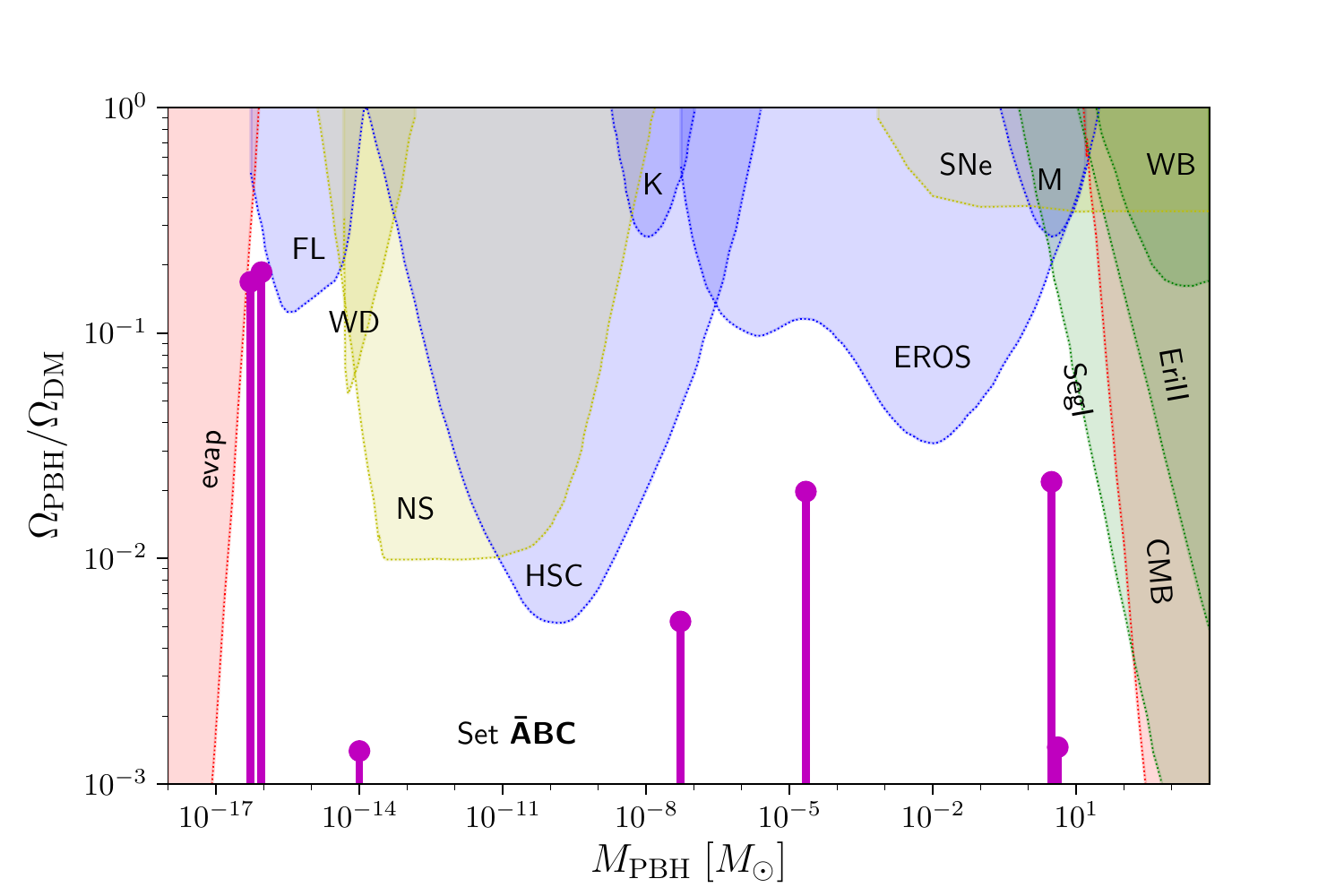}
\caption{The semi-analytical optimum mass function for four sets of constraints. Constraint functions for monochromatic mass functions are shown in blue (\cset{A}), red (\cset{\bar A}), green (\cset{B}), and yellow (\cset{C}). Vertical lines denote the locations of Dirac deltas in the semi-analytical optimum mass function, with height indicating the weight given to each one. The labeled constraints are from BH evaporation (\texttt{evap}, \cite{evap_Carr2010}), GRB femtolensing observations (\texttt{FL}, \cite{fl_Barnacka2012}), white dwarf explosions (\texttt{WD}, \cite{constraint_WD}), Hyper Suprime-Cam (\texttt{HSC}, \cite{hsc_Niikura2017}), Kepler (\texttt{K}, \cite{kepler_Griest2014}), EROS-II (\texttt{EROS}, \cite{constraint_EROS}), supernova lensing (\texttt{SNe}, \cite{constraint_SNe}), MACHO (\texttt{MACHO}, \cite{MACHO_Alcock2001}), Segue I dynamics (\texttt{SegI}, \cite{Seg_Koushiappas2017}), Eridanus II dynamics (\texttt{EriII}, \cite{Eri_Brandt2016}), wide binary dynamics (\texttt{WB}, \cite{WB_Monroy-Rodriguez2014}), and CMB observables (\texttt{CMB}, \cite{Planck_Ali-Haimoud2017,Carr2017}).
}
\label{fig:optimal-mass-functions}
\end{figure*}

We determine optimal mass functions for sets \cset{A}, \cset{\bar A}, and all of their combinations with sets \cset{B} and \cset{C}. The results are summarized in~\cref{tab:results} and illustrated in~\cref{fig:optimal-mass-functions}. We do not include cosmological constraints on the total matter density, so these values of $\fbound$ may exceed 1. In particular, note that all combinations containing \cset{A} have $\fbound>1$, while all combinations containing \cset{\bar A} and \cset{C} have $\fbound<1$. The set \cset{\bar A} on its own has marginal status if only monochromatic mass functions are considered, but clearly $\fbound>1$ in this case. With the constraints we consider in this work, $\fpbh=1$ is always allowed when using the less stringent set \cset{A}, regardless of additional constraints.
\begin{table}\centering
\begin{tabular}{lccccc}
\hline
& $\ftotmono$ & $\fbound$ & $\fgw$ & $\sigma[\psi]/M_\odot$ & $\left\langle M/M_\odot\right\rangle$
\\
\hline
\cset{A} & 27.17 & 27.25 & 2.580 & 2.259 & 31.09
\\
\cset{AB} & 1.372 & 1.965 & 5.139 & 0.162 & 0.009
\\
\cset{AC} & 1.371 & 1.443 & 0.566 & 7.294 & 1.807
\\
\cset{ABC} & 1.371 & 1.402 & 2.936 & 0.220 & 0.015
\\
\cset{\bar A} & 0.991 & 1.502 & 2.171 & 4.827 & 1.492
\\
\cset{\bar AB} & 0.991 & 1.437 & 11.07 & 0.221 & 0.017
\\
\cset{\bar AC} & 0.330 & 0.484 & 0.364 & 7.963 & 5.430
\\
\cset{\bar ABC} & 0.330 & 0.405 & 0.982 & 0.741 & 0.182
\\\hline
\end{tabular}
\caption{Optimal mass function properties for each of several sets of constraints. The column $\ftotmono$ gives the maximum DM fraction allowed for a monochromatic mass function, and the column $\fbound$ gives the maximum DM fraction across all functional forms. The column $\fgw$ gives the maximum DM fraction obtained by scaling the semi-analytical optimum while remaining consistent with gravitational wave constraints (see~\cref{sec:gravitational-waves}). Also given here are the mean PBH mass and the standard deviation for the semi-analytical optimum mass function.}
\label{tab:results}
\end{table}

\section{Prospects for gravitational wave constraints}
\label{sec:gravitational-waves}
Gravitational wave observables are the major exception to the rule that measured quantities are linear in the PBH mass function in each mass bin. There are several methods by which gravitational waves might constrain the primordial black hole population. In principle, the simplest constraint arises from present-day measurements of the black hole binary (BHB) merger rate, but this is weak for two reasons: first, it is difficult to distinguish primordial black holes from astrophysical black holes, for which a variety of additional physical mechanisms may affect the merger rate; and second, the observed merger rate is sufficiently uncertain as to be compatible with a wide range of PBH dark matter models \citep{Bird2016,Garcia-Bellido2017}.

An alternative method is to search for the stochastic gravitational wave background from primordial density fluctuations associated with inflationary production mechanisms \citep{Saito2010}. However, such constraints are only effective within the context of this class of formation models, and within such a limited scope, our level of generality is excessive. Here it is sufficient to consider the mass functions that can be reasonably produced by such formation mechanisms, and constraints for such mass functions have been treated elsewhere in the literature.

A third technique is to search for the stochastic gravitational wave background due to BHB mergers throughout cosmic history \citep{Ioka1999}. While such an approach may ultimately produce strong constraints, there remains a great deal of uncertainty in modeling the merger rate, particularly for extended mass functions. This problem has only recently been treated in the literature \citep{Clesse2017,Chen2018}, and the resulting constraints may not be robust. Still, it is useful to estimate these constraints, even imprecisely, in order to determine their relevance in the case of our semi-analytical optimum mass functions.

We now consider constraints from the non-detection of a stochastic background of gravitational waves from BHBs throughout cosmic history. This background is qualitatively different from all other observables considered in this work, since it has complicated non-linear dependence on the mass function. This means that determining constraints on a general mass function is non-trivial. In particular, one needs to include the higher order terms in the expansion of~\cref{eq:psi-observable}, and the analogue of~\cref{eq:observable-kernel} is then
\begin{equation}
\sum_{n=1}^\infty K_n(M_0,\dotsc,M_0)=\frac{A[\psimono]-A_0}{\fmaxmono(M_0)}
\end{equation}
which does not constrain off-diagonal values of the kernels $K_n$. Thus, gravitational wave constraints on the parameter space of monochromatic mass functions are insufficient to determine constraints on an extended mass function, even when the functional form is specified. This reflects the fact that gravitational wave constraints on extended mass functions are inherently model-dependent, in that one must determine the contribution to the background from binaries whose partners have unequal masses.

The results of~\cite{Clesse2017} provide a simple method for estimating the stochastic gravitational wave background given a particular mass function, which we now review. The observable characteristic strain amplitude is given by
\begin{equation}
h_c^2(\ngw)=\frac{4A_1}{3\pi^{1/3}\left(\log10\right)^2}\left(\frac{GM_\odot}{c^2}\right)^{5/3}\left(\frac{\ngw}{c}\right)^{-4/3}\int\frac{\du m_1}{m_1}\frac{\du m_2}{m_2}\,\tau_{\mathrm{merge}}\left(m_1,m_2\right)\mathcal M_c^{5/3}
\end{equation}
where $A_1\simeq0.7642H_0^{-1}$ is a cosmology-dependent constant, $\ngw$ is the gravitational wave frequency, $\tau_{\mathrm{merge}}$ is the mass-dependent binary merger rate per unit volume, and $\mathcal M_c\equiv(m_1m_2)^{3/5}(m_1+m_2)^{-1/5}$ is the chirp mass. The merger rate is determined by consideration of the capture rate for formation of PBH binaries, given in the Newtonian approximation by
\begin{equation}
\tau_{\mathrm{capture}}(m_1,m_2)=2\pi n_{\mathrm{PBH}}(m_1)v_{\mathrm{PBH}}\left(\frac{85\pi}{6\sqrt2}\right)^{2/7}\frac{G^2(m_1+m_2)^{10/7}(m_1m_2)^{2/7}}{(v_{\mathrm{rel}}/c)^{18/7}c^4}
\end{equation}
where $n_{\mathrm{PBH}}$ is the local number density of PBH, $v_{\mathrm{PBH}}$ is the characteristic velocity of a single black hole, and $v_{\mathrm{rel}}\equiv\sqrt2v_{\mathrm{PBH}}$ is the characteristic relative velocity of two black holes. The local number density of PBH of mass $m$ is parametrized as $n_{\mathrm{PBH}}=\delta_{\mathrm{PBH}}\rho_{\mathrm{PBH}}(m)/m$, where $\rho_{\mathrm{PBH}}(m)$ is the cosmological average density of PBH of mass $m$ and $\delta_{\mathrm{PBH}}$ is the local density contrast of PBH. In particular, in our notation, this number density is given by
\begin{equation}
n_{\mathrm{PBH}}(m)=\delta_{\mathrm{PBH}}\frac{\Omega_M\rho_c\psi(m)}{m}
\end{equation}
Estimates for $\delta_{\mathrm{PBH}}$ range from $10^6$ to $10^{10}$. In order to estimate conservative constraints, we henceforth take the relatively low value $\delta_{\mathrm{PBH}}=10^7$.

In terms of the capture rate, the merger rate per unit volume is $\tau_{\mathrm{merge}}(m_1,m_2)=(\fpbh/\delta_{\mathrm{PBH}})\tau_{\mathrm{capture}}(m_1,m_2)n_{\mathrm{PBH}}(m_2)$. Thus the strain amplitude is given by
\begin{equation}
\label{eq:strain-amplitude}
h_c^2[\psi](\ngw)=\left(\frac{\ngw}{c}\right)^{-4/3}C\int\du m_1\dd m_2\,\frac{\psi(m_1)}{m_1}\frac{\psi(m_2)}{m_2}\left(m_1m_2\right)^{2/7}\left(m_1+m_2\right)^{23/21}
\end{equation}
where $C$ is a cosmology-dependent factor given by
\begin{equation}
C=\frac{2\pi^{20/21}170^{2/7}c^{1/7}A_1}{3^{9/7}(\log10)^2}\left(\frac{GM_\odot}{c^2}\right)^{5/3}\frac{\rho_M^2\delta_{\mathrm{PBH}}\fpbh}{(v_{\mathrm{PBH}}/c)^{11/7}}.
\end{equation}
In particular, the dependence of $h_c^2(\ngw)$ on $\psi$ admits a simple expansion of the form of~\cref{eq:psi-observable}. If $\fpbh$ is fixed independently of $\psi$, then the only non-vanishing term has the form $\int\du m_1\dd m_2\,\psi(m_1)\psi(m_2)K_2(m_1,m_2)$, with
\begin{equation}
K_2(m_1,m_2)=C\left(\frac{\ngw}{c}\right)^{-4/3}\frac{(m_1+m_2)^{23/21}}{(m_1m_2)^{5/7}}.
\end{equation}
Note that $K_2$ varies with $\ngw$, reflecting the fact that measurements of $h_c(\ngw)$ at each frequency $\ngw$ are independent. Thus, if the instrument used has sufficiently high frequency resolution, a large number of independent constraining observables can be measured, meaning that the maximizing mass function need not resemble a linear combination of a small number of monochromatic mass functions. In this case, direct numerical methods are necessary to determine the maximum density of PBH.

It would be inappropriate to perform a full numerical optimization within our framework, since there are considerable theoretical uncertainties in the determination of merger rates. However, we can estimate the potential impact of gravitational wave constraints by checking compatibility of our optimal mass functions with existing gravitational wave observations. This serves to indicate the potential for future modeling work to constrain the PBH population: if the semi-analytical optimum for a given set of constraints is already consistent with gravitational wave observations as well, then we can predict that the detailed inclusion of this additional constraint will have a minimal impact on $\fbound$.

Current aLIGO bounds on $h_c(\ngw)$ are strongest at $\ngw\simeq\SI{100}{\hertz}$, and we represent the current limit by $h_c(\SI{100}{\hertz})\lesssim10^{-22}$. Given a functional form for the mass function $\psi$, we can compute the maximum $\fpbh$ for which $h_c[\psi](\SI{100}{\hertz})$ satisfies this bound. We denote this maximum by $\fgw$, and the values of $\fgw$ for our semi-analytical optima are shown in~\cref{tab:results}. Of the sets of constraints we consider, only sets \cset{AC} and \cset{\bar AC} have $\fgw$ significantly less than one. This is to be expected, since the maximizing mass function in both of these cases has a large variance: \cite{Clesse2017} show that the gravitational wave background is strongly enhanced as the variance is increased. In the case of set \cset{\bar AC}, we have $\fgw\simeq\fbound<1$, meaning that the overall maximum is minimally impacted by gravitational wave constraints. In particular, $\fpbh=1$ is ruled out regardless.

Only set \cset{AC} has $\fgw<1<\fbound$, which, in general, is difficult to interpret: in principle, there may be a different form for the mass function which relaxes gravitational wave constraints, retains $\fpbh\geq1$, and remains consistent with the other constraints we consider in this work. A simple way to check this is to consider the maximizing monochromatic mass function, for which gravitational wave constraints should be relaxed compared with the high-variance semi-analytical optimum. Indeed, there is a monochromatic mass function which satisfies all non-gravitational constraints in set \cset{AC} with $\fpbh=1.371$, and we compute $\fgw>10$ for this mass function. Thus, while it may not be possible to attain $\fbound$ in this case without violating gravitational wave constraints, $\fpbh=1$ clearly remains allowed. As such, the addition of gravitational wave constraints does not change the overall status of the PBH dark matter paradigm for any of the constraint sets we consider. This reflects both the current status of observations and the large uncertainties in modeling the background. However, future experiments are expected to improve limits on $h_c(\ngw)$ by 2--4 orders of magnitude, which would be sufficient to rule out all of the mass functions represented in~\cref{tab:results} even under fairly conservative assumptions.

\section{Discussion}
\label{sec:discussion}
With the maximization procedure introduced in~\cref{sec:general}, it is simple to determine the maximum PBH density consistent with constraints. We stress that this is a bound that applies for mass functions of all forms. Thus, given a set of observational constraints, we can determine a model-independent bound on the density of PBH.

Our results quantify, for the first time, the risks of using monochromatic mass functions to assess the overall status of the PBH dark matter paradigm. So long as one window in the constraint functions is much less constrained than all others, the difference between $\fbound$ and $\ftotmono$ is generally very small. Set \cset{A} is a clear example of such a case, and the correction is of order 0.1\%. On the other hand, if PBH are constrained to a similar extent in multiple windows, the correction can be large. The most dramatic example is provided by set \cset{\bar A}, for which $\fbound$ is larger than $\ftotmono$ by $\sim50\%$. We conclude that, at worst, the bound on the total PBH density is related to the monochromatic bound by an $O(1)$ factor.

The optimal mass functions themselves (\cref{fig:optimal-mass-functions}) do not correspond to any well-motivated production scenario that we are aware of, and we certainly do not claim that the maximal density can be attained by producing PBH monochromatically at a discrete collection of masses spanning 15 orders of magnitude. Instead, the panels of~\cref{fig:optimal-mass-functions} should be interpreted as a tool to relate monochromatic constraint functions to their impact on the allowed total density of PBH. In particular, an immediate and non-trivial conclusion that can be drawn from the figures is that the addition of any new constraint which does not overlap the peaks of the optimal mass function will not reduce $\fbound$.

Further, the functional form of the optimal mass function clarifies the dependence of constraints on the variance of the mass function. In the single-constraint case, we showed that an extended mass function never outperforms the optimal monochromatic mass function. Indeed, in this case, increasing the variance of a narrow mass function will only relax constraints if $\fmaxmono$ is concave-up in the mass range of interest, i.e., if the monochromatic mass function under consideration is not the optimal one. When multiple constraints are considered, the relationship between the variance of the mass function and the allowed density is less obvious. Our semi-analytical optimum mass functions all exhibit some non-zero spread, and they definitively allow higher PBH densities than any zero-variance (i.e., monochromatic) mass function. However, extending a monochromatic mass function only slightly, without overlapping additional points of the semi-analytical optimum mass function, is not useful for relaxing constraints. In this respect, our findings are consistent with those of~\cite{Carr2017,Bellomo2017}.

The most substantial differences in $\fbound$ arise from differences between \cset{A} and \cset{\bar A}. Set \cset{\bar A} contains more stringent forms of constraints from CMB anisotropy and PBH evaporation. The CMB constraint is strongly dependent on modeling poorly-understood accretion processes. Both versions of the constraint used in this work are drawn from~\cite{Planck_Ali-Haimoud2017}: the version in set \cset{A} is obtained by considering only collisional ionization of the accreted gas, while the version in set \cset{\bar A} is obtained by including photoionization as well. The evaporation constraint is sensitive to uncertainties in the spectrum of extragalactic background radiation. We adopt the extreme cases considered by \cite{Carr2017}, with the relaxed form contained in set \cset{A} and the more stringent form in set \cset{\bar A}.

\subsection{Relative impact of constraints}
The values of $\fbound$ in~\cref{tab:results} demonstrate that the present observational status of PBH dark matter is strongly dependent on the constraints adopted. However, to rule out $\fpbh=1$, it is necessary to both take the more stringent constraints \cset{\bar A} in place of \cset A, and to include at least one of the constraints from set \cset{C}: supernova microlensing \citep{constraint_SNe}, neutron star capture \citep{constraint_NS}, and white dwarf explosions \cite{constraint_WD}.

The supernova microlensing constraint is the most recent of those we consider, and its robustness is the subject of ongoing discussion in the literature \citep[see e.g.][]{Garcia-Bellido2017}. We note that this constraint is dominant in the LIGO window only when dynamical constraints from set \cset{B} are neglected, so the addition of this constraint alone to set \cset{AB} or \cset{\bar AB} will have a small impact on $\fbound$. The constraint from neutron star capture is also subject to astrophysical uncertainties, since it is dependent on the dark matter density in the cores of galactic clusters \citep{constraint_NS}. We consider the relatively restrictive constraint obtained by taking $\rho_{\mathrm{DM}}=\SI[parse-numbers=false]{10^4}{\giga\electronvolt\cm^{-3}}$. The strength of the constraint scales linearly with $\rho_{\mathrm{DM}}$, and more conservative estimates take $\rho_{\mathrm{DM}}$ smaller by an order of magnitude or more. However, this constraint is most effective in a window shared with constraints from white dwarf explosions, so even if one of the two is subject to substantial uncertainties, the effect of set \cset{C} on $\fbound$ remains large.

The form of the optimal mass function allows us to rapidly identify the potential impacts of prospective constraints from future observations. For instance, constraints on intermediate-mass black holes with $M\gtrsim10^2\,M_\odot$ are already strong enough that our semi-analytical optimal mass functions are negligibly small throughout this region. Thus, the identification of additional dynamical systems that might tighten constraints in this region will not affect the overall bound on the PBH density at a level greater than one part in $10^4$. On the other hand, GRB femtolensing limits lie in a mass range where some of the semi-analytical optima have a large peak, and strengthening these constraints will have an immediate impact on the overall bound. In particular, upcoming Fermi GRB observations are expected to substantially strengthen constraints in this window, improving by a factor of five after 10 years of operation \cite{Barnacka2013}.
% This remark is found at the bottom of page 158.
These results may ultimately rule out the PBH dark matter paradigm, with the exception of non-evaporating Planck-mass relics.

Constraints from gravitational wave observations are a special case, as they do not admit the linear interpretation that we take for the impact of extended mass functions on other observables. Further, the strain amplitude at each frequency is sensitive to PBH in a wide range of masses. Thus, it is not trivial to predict the effect of future gravitational wave constraints on our overall bound without direct numerical optimization. However, constraints from LISA and DECIGO \citep[see e.g.][]{Moore2015} will eventually be capable of ruling out all of our semi-analytical optima, potentially lowering the upper bounds we set in this work.

\section{Conclusions}
\label{sec:conclusions}
We have found the form of the mass function which maximizes the PBH density subject to observational constraints, and we have used this to calculate an upper bound on the fraction of dark matter in PBH. Depending on the constraints adopted, we find $\fbound$ as large as 27.25 (set \cset A) or as small as 0.405 (set \cset{\bar ABC}). The scenario in which all dark matter is composed of PBH is ruled out by stringent limits from evaporation and Planck if combined with the constraints from white dwarf explosions, neutron star capture and SNe lensing (set \cset{C}). However, if relaxed constraints from evaporation and Planck are adopted, PBH dark matter is not ruled out by the addition of any other constraints we consider in this work. Estimated gravitational wave constraints do not affect these conclusions at the sensitivity of current instruments.

Our method provides a fast and robust technique to determine the total allowed density of PBH given a set of constraints ($\fbound$), independent of the form of the PBH mass function. The optimal mass function itself allows an easy test of the impact of additional constraints on $\fbound$. While the optimal mass function is not exactly monochromatic, it is very nearly so for realistic constraints. The optimal mass function corresponding to each set of constraints we consider is approximately monochromatic, with additional components scaling the total allowed fraction by no more than an $O(1)$ factor. Our results explain the findings of \cite{Carr2017,Bellomo2017} that extended mass functions are generally more strongly constrained than monochromatic mass functions, and confirm that the monochromatic maximum density $\fmono$ is a good approximation of the allowed density across all mass functions.

\appendix

\section{Numerical validation}
\label{sec:numerical}

Given a set of constraints, it is also possible to use numerical methods to find a mass function which maximizes the PBH density. There are significant caveats to such an approach. Most importantly, a maximization algorithm may converge to a local optimum rather than a global optimum. Additionally, computational costs may render numerical approaches impractical unless the functions involved are discretized sparsely. Even so, numerical optimization can be used to validate our analytical results: if the same set of masses is used for discretization, then the numerical result should never reach a greater normalized mass (cf.~\cref{eq:multi-normalized-mass}) than that of our corresponding semi-analytical result. Numerical methods can also be used to check that our semi-analytical optimum is a stationary point of the normalized mass functional.

\subsection{Direct validation}
\label{sec:numerical-direct}
We implement these validation steps using a simple Monte Carlo algorithm, as follows: we begin with an initial mass function of the form $\psi_0(M)\propto M^{-1}$, which assigns equal PBH density to each log-spaced mass bin. We then perturb the value of $\psi_0$ in a random bin $k$ by a value selected from a Gaussian distribution with mean 0 and variance $\sigma^2\psi_0(M_k)^2$, where $\sigma$ is a parameter of the maximization. We denote the resulting mass function by $\psi_1(M)$. If $\psi_1(M_k)\geq0$ and $\mathcal M[\psi_1]>\mathcal M[\psi_0]$, we accept the step, replace $\psi_0$ by $\psi_1$, and repeat. For simplicity, we do not accept any steps which reduce the normalized mass. This is not necessary in order to test whether our semi-analytical optimum mass function is a stationary point. We also reject steps which increase the normalized mass by less than $10^{-10}$ to avoid exceeding the numerical precision of the semi-analytical result.

In order to make the problem numerically tractable, we use only $10^2$ log-spaced mass bins. This discretization is different from the one used in~\cref{tab:results}, and it does not capture sharp features of the constraints. Consequently, in order to compare the numerical results with semi-analytical results, we regenerate the semi-analytical mass function with the same discretization. Note that this affects both the form of the optimal mass function and the calculated $\fbound$.

\begin{figure*}\centering
\includegraphics{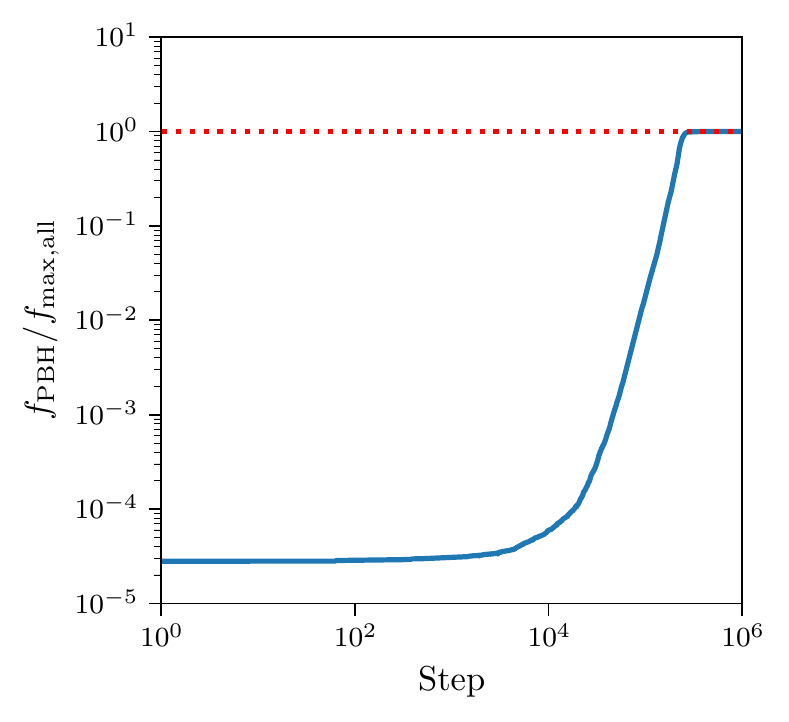}
\hspace{-0.2cm}
\includegraphics{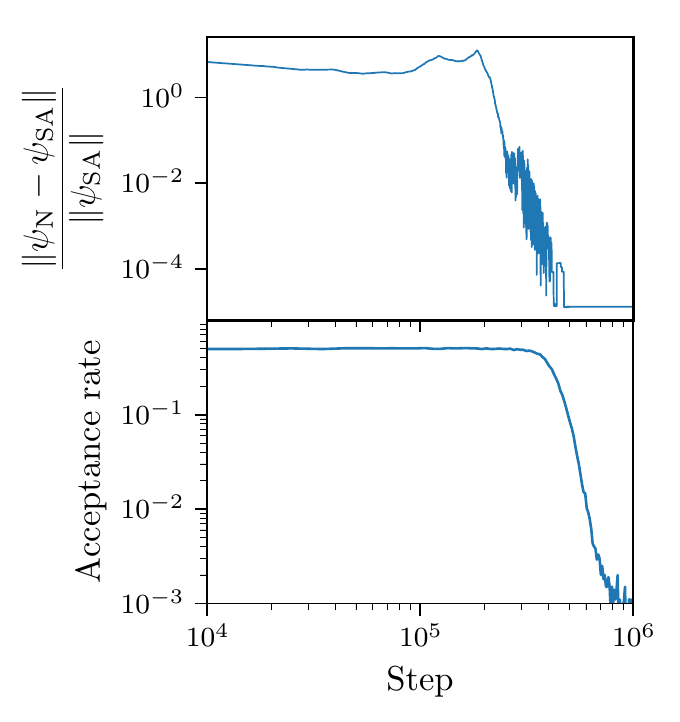}
\caption{
Left: $\fpbh$ attained in each step during numerical maximization, shown as a fraction of the semi-analytical $\fbound$. The dashed red line indicates $\fpbh=\fbound$. Right top: $L^2$ norm of the difference between $\psi_{\mathrm{N}}$ (numerical) and $\psi_{\mathrm{SA}}$ (semi-analytical) mass functions for each step, shown as a fraction of $\norm{\psi_{\mathrm{SA}}}$. In computing the norm, $\psi_{\mathrm{SA}}$ is treated as a step function on the mass bins. Right bottom: acceptance rate in bins of $10^4$ steps.
}
\label{fig:numerical-convergence}
\end{figure*}

We implement the numerical optimization with $\sigma=10^{-2}$. In what follows, we denote the numerical mass function by $\psi_{\mathrm{N}}$, and the semi-analytical optimum by $\psi_{\mathrm{SA}}$. The left-hand side of \cref{fig:numerical-convergence} shows $\fpbh$ for the numerical mass function at each step as a fraction of the semi-analytical $\fbound$. The numerical $\fpbh$ converges to $\fbound$ and immediately stabilizes, and in particular, in no step does $\fpbh$ exceed $\fbound$.

In principle, $\psi_{\mathrm{N}}$ need not converge to $\psi_{\mathrm{SA}}$ even given that $\fpbh$ converges to $\fbound$, since the mass function with maximal density is not necessarily unique. However, in the top-right panel of~\cref{fig:numerical-convergence}, we show that $\psi_{\mathrm{N}}$ tends to $\psi_{\mathrm{SA}}$ in the $L^2$ norm. To compute this distance consistently, we treat the Dirac deltas of $\psi_{\mathrm{SA}}$ as constant functions in their respective bins. As an additional test of convergence, we compute the acceptance rate, i.e., the fraction of steps which are accepted, during each window of $10^4$ iterations. The acceptance rate vanishes as $\psi_{\mathrm{N}}$ approaches $\psi_{\mathrm{SA}}$, which further demonstrates that $\psi_{\mathrm{SA}}$ is a stationary point of the normalized mass.

\begin{figure*}\centering
\includegraphics{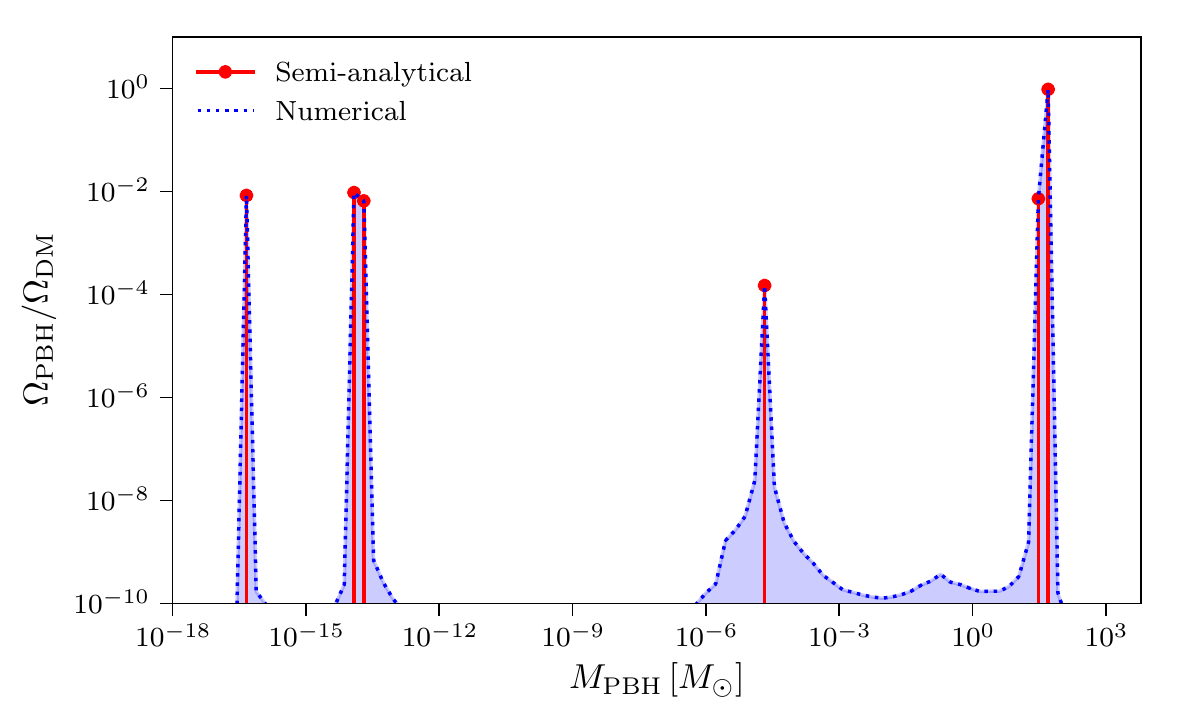}
\caption{Blue: numerically-optimized mass function $\psi_{\mathrm{N}}$ after $10^6$ steps. Red: semi-analytical optimum $\psi_{\mathrm{SA}}$. Each curve shows the integral of the mass function in each bin, i.e., the total contribution of that bin to $\fpbh$.}
\label{fig:numerical-comparison}
\end{figure*}

The numerical and semi-analytical mass functions are shown in~\cref{fig:numerical-comparison}. In order to compare Dirac deltas with the smooth mass function $\psi_{\mathrm{N}}$, the figure shows the integral of the mass function in each bin rather than $\psi_{\mathrm{N}}$ and $\psi_{\mathrm{SA}}$ themselves. It is clear that in this case, the numerical algorithm converges to the semi-analytical optimum. We have established via analytical arguments that this is not simply a local optimum, but indeed the global maximum of the normalized mass.

\subsection{Sensitivity to the constraint prescription}
\label{sec:prescription-sensitivity}
Our analytical work is based on the prescription of~\cite{Carr2017} for evaluating constraints on extended mass functions. Since other prescriptions have been considered in the literature, it is important to determine the robustness of our results to variations on the constraint prescription. Assessing this analytically is intractable, as it requires the development of independent analytical frameworks for even slight modifications. However, numerical methods allow for a comparison of the bounds we obtain analytically with those that would result from any other specified prescription. We thus perform numerical optimization under the prescriptions of~\cite{Bellomo2017} and~\cite{Carr2016}, and compare these with our semi-analytical results.

The situation is particularly simple for the constraint prescription of~\cite{Bellomo2017}: the major difference is that a mass function is allowed if it is allowed according to each individual constraint, rather than according to their statistical combination. Thus, the normalized mass $\mathcal M[\psi]$ is replaced by
\begin{equation}
\widehat{\mathcal M}[\psi]\equiv\frac{\int\du M\,\psi(M)}{\max_{j=1,\dotsc,N}\mathcal C_j[\psi]}.
\end{equation}
It is straightforward to implement numerical optimization with respect to $\widehat{\mathcal M}[\psi]$ in place of $\mathcal M[\psi]$. For the case shown in~\cref{fig:numerical-comparison}, these two numerical maxima agree to within 1\%.

We also implement the constraint procedure of~\cite{Carr2016}, for which the constraints are treated as step functions on a set of mass bins. There is no universal prescription for the size of the bins across all constraints, but they should be chosen small enough that the minimum of each constraint function is not very different from its maximum within any single bin. Altogether, the procedure is as follows:
\begin{enumerate}
\item The mass range is divided into bins $I_1,\dotsc,I_n$.
\item In each bin, the dominant constraint is identified. If a bin captures the transition between two dominant constraints, the bin is subdivided at the transition point.
\item We evaluate the constraint of~\cref{eq:discrete-constraint}, considering only the dominant constraint in each bin. We treat $\psi$ as a smooth function, using a more refined set of bins for its numerical representation.
\end{enumerate}
As in~\cref{sec:numerical-direct}, we take a relatively coarse binning of the mass range for numerical testing purposes. We have evaluated the maxima attained for a range of different constraint bin counts, and we find that our determination of $\fbound$ is robust to changes in binning at the 10\% level.

We note as well that prescriptions of this kind have been criticized in the literature for the fact that it is not trivial to determine the range of validity of the individual constraints, and hence to determine the limits of integration for each constraint curve in~\cref{eq:general-constraint}. This is indeed a concern when individual constraints are considered, and it introduces significant potential uncertainty in cases where there are masses for which all constraints are at the edge of their range of validity. In these scenarios---for instance, with the constraints of set \cset{A}---there are effectively ``windows'' in the constraints that make any upper bound on the density of PBH quite uncertain. However, in the other cases that we consider, the dominating constraints in a given mass range generally intersect well within their respective ranges of validity. Thus, modifying the limits of integration does not substantially impact our results.

\bibliography{maximizing_pbh_mass}

\end{document}